\documentclass[english,aps, prl, twocolumn]{revtex4-1}
\usepackage[T1]{fontenc}
\usepackage[latin9]{inputenc}
\usepackage{graphicx}

\makeatletter

\providecommand{\tabularnewline}{\\}

\makeatother

\usepackage{babel}
\begin{document}
\title{Sample-efficient benchmarking of multi-photon interference on a boson
sampler in the sparse regime}
\author{Jelmer J. Renema\textsuperscript{1)}, Hui Wang\textsuperscript{2,3)},
Jian Qin\textsuperscript{2,3)}, Xiang You\textsuperscript{2,3)},
Chaoyang Lu\textsuperscript{2,3)}, Jianwei Pan\textsuperscript{2,3)}}
\affiliation{1) Mesa+ Institute for Nanotechnology, University of Twente, PO Box
217, 7500 AE Enschede, The Netherlands.~\linebreak{}
2) Hefei National Laboratory for Physical Sciences at Microscale and
Department of Modern Physics, University of Science and Technology
of China, Hefei, Anhui 230026, China. ~\linebreak{}
3) CAS Centre for Excellence and Synergetic Innovation Centre in Quantum
Information and Quantum Physics, University of Science and Technology
of China, Hefei, Anhui 230026, China}
\begin{abstract}
Verification of a quantum advantage in the presence of noise is a
key open problem in the study of near-term quantum devices. In this
work, we show how to assess the quality of photonic interference in
a linear optical quantum device (boson sampler) by using a maximum
likelihood method to measure the strength at which various noise sources
are present in the experiment. This allows us to use a sparse set
of samples to test whether a given boson sampling experiment meets
known upper bounds on the level of noise permissible to demonstrate
a quantum advantage. Furthermore, this method allows us monitor the
evolution of noise in real time, creating a valuable diagnostic tool.
Finally, we observe that sources of noise in the experiment compound,
meaning that the observed value of the mutual photon indistinguishability,
which is the main imperfection in our study, is an effective value
taking into account all sources of error in the experiment. 
\end{abstract}
\maketitle

\section{Introduction}

With the increasing computational power of quantum information processing
devices \cite{Arute2019,Pino2020,Wang2019b}, it has become a pressing
problem how to demonstrate the computational advantage that a quantum
device has over a classical one (also known as 'quantum supremacy').
A key issue is \emph{verification}, i.e. the problem of checking that
the quantum device is truly outperforming classical computer. Ideally,
a quantum advantage demonstration would be arranged so that verification
can be done efficiently with respect to the size of the quantum system.
For example, an efficiently verifiable quantum advantage can be demonstrated
on a universal, fault-tolerant computer by a fast solution to a problem
such as prime factoring, i.e. one whose output can be efficiently
checked, and for which a fast quantum algorithm is known but not a
classical one. If the quantum machine can solve the posed problem
much faster than a classical computer, this would constitute very
strong evidence of the quantum nature of the device. 

However, for the forseeable future, the main devices of experimental
interest are noisy intermediate-scale quantum devices (NISQ), which
are small-to-medium sized devices without quantum error correction.
These devices usually work by solving a sampling problem over some
probability distribution more efficiently than is believed to be possible
classically \cite{Preskill2018,Harrow2017}. In photonics, this takes
the form of sampling over the output distribution of interfering photons
in a linear optical network, a problem known as boson sampling (see
Figure 1) \cite{Aaronson2011}. For such devices, no problems are
known which fit all three requirements for efficient verification
(easy to solve on a NISQ device, hard to solve classically, easy to
verify). Informally stated, the issue is that it has not proven possible
to find some property to be hidden in the distribution which can be
efficiently tested for quantumly but not classically. Therefore, the
usual approach in proving quantum operation in such a device is a
brute force one, where the operation of the device is directly reconstructed
at a computational cost exponential in the system size. 

A further problem in the demonstration of a quantum advantage using
NISQ devices is the influence of noise, i.e. any unwanted physical
effect which degrades the performance of the device. For NISQ devices,
the physical picture is that noise pushes the distribution over which
the device samples closer to the distribution associated with classical
operation of the device \cite{Boixo2018}. For quantum advantage demonstrations
based on random circuit sampling \cite{Arute2019}, it was recently
proposed that if the noise pushes the system towards the uniform distribution,
any level of noise which doesn't completely revert the device to sampling
over a classical distribution is sufficient to maintain a quantum
advantage \cite{Aaronson2019}. However, this result relies on strong
complexity-theoretic assumptions regarding the classical hardness
of identifying properties of the probability distribution over which
the device is sampling. 

For boson sampling, in contrast, pseudoprobability distributions are
known which approximate the sampling distribution in the presence
of noise \cite{Renema2018a,Renema2018b,Shchesnovich2019a,Shchesnovich2019b,Moylett2019,Renema2019},
and from which it is believed to be possible to sample efficiently.
The distance between these distributions and the output distribution
of the device depends on the level of noise, and for sufficiently
high levels of noise the two distributions are close, ruling out a
quantum advantage in that regime.

However, the application of such classicality thresholds requires
knowledge of which noise sources are present in a boson sampler, and
at what strength. The main sources of noise in a boson sampler are
photon loss \cite{Aaronsonweb2} \footnote{While photon loss can be removed by postselecing on those events where
all photons make it through the optical system, this comes the cost
of an exponential overhead in the time required to produce a sample}, and partial distinguishability \cite{Tichypartial}. So far, the
level of noise in boson sampling experiments was mostly inferred by
measuring the complete output distribution and comparing it to the
ideal (i.e. noiseless) output distribution, where the experiment is
declared a success if the distance between these is sufficiently low.
However, this approach is inherently non-scalable in the number of
measurements required \cite{Aaronson2011} since the probability of
observing any single outcome decreases exponentially with the number
of photons. The other approach was to compare the hypothesis of fully
distinguishable photons with that of fully indistinguishable ones
\cite{Bentivegna2014}, which does not capture the full range of possible
effects in the device \cite{Tichypartial}. This issue recently became
pressing with the demonstration of the first boson sampler operating
in the sparse regime \cite{Wang2019b}, i.e. where the size of the
Hilbert space makes measuring the output distribution infeasible,
and only a sparse set of samples can be collected. This raises the
question of how to assess the quality of quantum interference in such
a system.

In this work, we show that all information required to assess the
quality of quantum interference in a boson sampler is contained in
just a sparse set of samples. We use a maximum likelihood method to
infer the relevant properties of the photons, in particular their
mutual indistinguishability. We apply this method to a sparse set
of samples produced on the experimental apparatus reported in \cite{Wang2019b}.
There is a discrepancy between the measured indistinguishability in
the sampler and that inferred from independent measurements, which
we explain by demonstrating that all other imperfections result in
a decrease of the indistinguishability. This value must therefore
be interpreted as an effective value taking into account many sources
of noise. Using this fact, we can construct an error budget for a
boson sampler. We also show how to use this method to monitor the
quality of photonic interference in real time, showing its use as
a diagnostic tool. Our method is essentially a non-trivial extension
of the Hong-Ou-Mandel interference \cite{Hong1987} to measure the
mutual overlap of photons in a highly complex multi-photon, multi-mode
case.

Together with the results on the permissible level of noise in a boson
sampler \cite{Renema2018a,Renema2018b,Shchesnovich2019a,Shchesnovich2019b,Moylett2019,Renema2019},
these results solve the problem of testing whether a given boson sampling
experiment passes the known requirements to exibit a quantum advantage
without any additional characterization experiments besides measuring
the transmission matrix of the system, which can be done efficiently
\cite{Laing2012}. Our method is efficient in the number of samples
required, but not efficient in the number of computations required,
as is expected for a NISQ quantum advantage verification protocol. 

Our work is structured as follows: we begin with a short overview
of the experimental apparatus on which the samples to which we apply
our method were produced. We then detail our maximum likelihood analysis
method. Finally, we conduct numerical simulations to show that the
effect of other imperfections is to decrease the degree of indistinguishability.

\begin{figure}
\includegraphics[width=7cm]{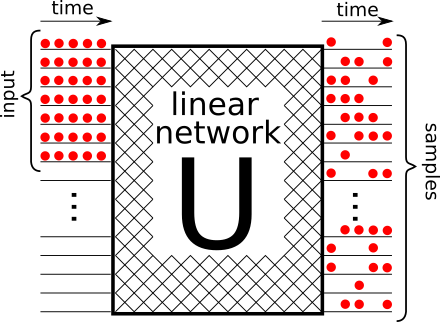}

\caption{A sketch of a boson sampler. A series of $n$ modes out of a large
linear interferometer are fed with single photons, and samples from
the resulting distribution are recorded at the output.}
\end{figure}

\section{experimental imperfections}

The experimental apparatus on which our dataset was generated consists
of an InAs/GaAs quantum-dot source operating at 893 nm pumped with
a pulsed laser with 1.2 nW intensity and a repetition rate of 76 MHz,
corresponding to a pulse energy of $1.6\times10^{-17}\ $J. This single-photon
source is then demultiplexed using a tree of 19 Pockels cells, and
fed into a fixed free-space interferometer with 60 optical modes,
where quantum interference occurs. Finally, detection is done by a
bank of superconducting single-photon detectors which are fiber-coupled
to these modes. Samples consist of the set of detectors which are
triggered in each run of the experiment, and are recorded using standard
correlation electronics. More details of the setup are provided in
\cite{Wang2019b}. 

The dataset under consideration here consists of lists of samples
from this device. For simplicity of the analysis, postselection was
used to focus on those samples where the number of photons incident
is equal to the number of detection events reported, thereby removing
the effect of photon loss from consideration. The number of detected
photons varied between $n=3$ and $n=7$, with the length of the list
of samples varying from 161 at $n=6$ to $3\times10^{4}$ at $n=3$.
Besides the samples, the dataset also contains the transmission matrix
$M$ of the interferometer, which was characterized independently.
A further restriction on the list of samples is given by the threshold
nature of the single-photon detectors combined with postselection,
namely that all samples must be \emph{collision-free} in order to
be registered, i.e. all $n$ photons must emerge from distinct output
modes. 

This experimental setup contains four imperfections of note, which
result in samples which do not correspond to ideal $n$-photon quantum
transmission in the interferometer. First: the generated photons are
not perfectly indistinguishable, which is a requirement for maximally
strong quantum interference. Second, the input states occasionally
contain additional noise photons. This results in samples where one
single photon is lost, but where this is compensated by one of the
modes containing two photons instead, preserving the total photon
number. Third, the detectors in the system produce dark counts (false
positives), resulting in events where one photon is lost and replaced
by a dark count. Finally, there is uncertainty in measuring the transmission
matrix $M$, resulting in differences between the actual and expected
interference patterns. 

\section{Photon distinguishability}

To measure the strength of these imperfections, we use a standard
maximum likelihood approach. We will begin by considering only indistinguishability,
adding in the other imperfections later. 

The intuition behind the maximum likelihood method is that if we have
some probability distribution $p$, which is a function of a set of
parameters $\theta,$ we can estimate $\theta$ from a list of samples
by the following expression: $\theta_{ml}=\mathrm{argmax_{\theta}}(\Lambda(\theta)),$
with $\Lambda(\theta)=\prod_{i}p_{i}(\theta)$ i.e. we must maximize
with respect to $\theta$ the total likelihood to find the observed
series of samples. Under relatively benign mathematical assumptions,
such as continuity of the likelihood function, this is an unbiased
and optimal way of estimating $\theta$.

\begin{figure}
\includegraphics[width=8.8cm]{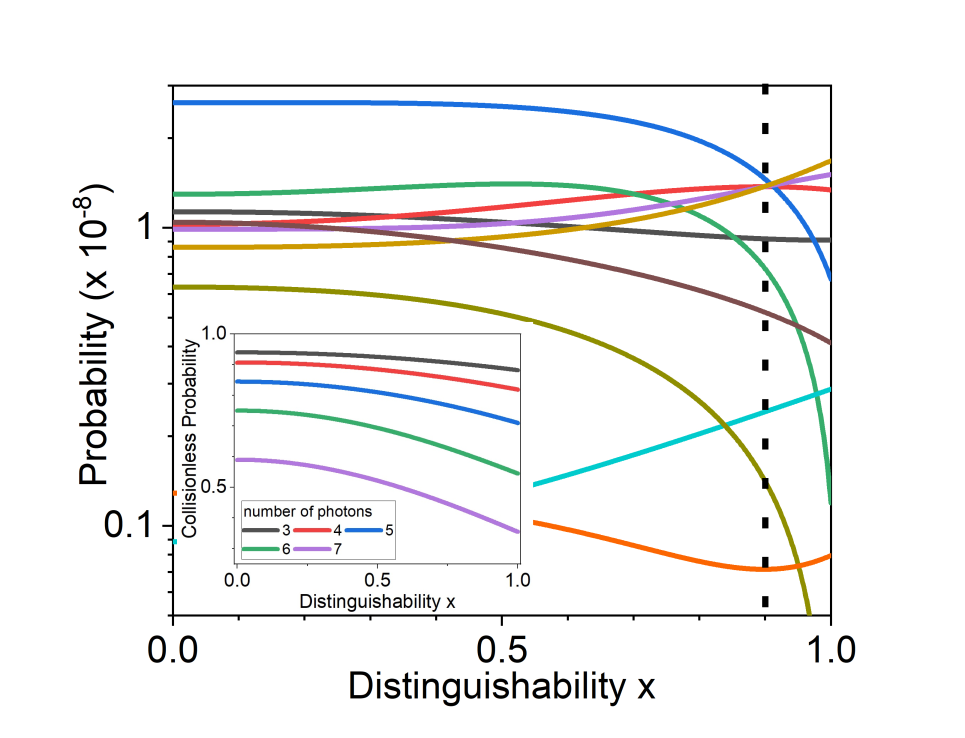}

\caption{The effect of partial photon distinguishability $x$ on the probability
for 10 randomly chosen output configurations\emph{. Inset}: Monte
Carlo estimate of the correction factor arising from the restriction
of our dataset to collisionless samples.}

\end{figure}

We will begin by focussing only on the effect of partial distinguishability.
In a boson sampler operating at some level of partial distinguishability,
the probability of a given outcome can be written as \cite{Shchesnovich2014,Shchesnovich2015a,Shchesnovich2015b}:
\begin{equation}
P(X)=\sum_{\sigma\in S_{n}}\left(\prod_{j}S_{j,\sigma_{j}}\right)\mathrm{Perm}(M\circ M_{\sigma,1}^{\dagger}),
\end{equation}
where $S$ is a matrix of distinguishabilities defined elementwise
as $S_{ij}=\langle x_{i}|x_{j}\rangle,$ $|x\rangle_{i}$ is the internal
wave function of the $i$-th photon, $\mathrm{Perm}$ is the permanent
function, $\sigma$ is a permutation of size $n,$ indices on matrices
denote permutation according to those indices, and $M$ is the submatrix
of $U$ connecting the modes containing an input photon to the outputs
of interest \cite{Scheel2008}, and $\circ$ is the elementwise product. 

It is this matrix $S$ which we are interested in estimating. Throughout
this work, we will parametrize $S$ elementwise as $S_{ij}=x+(1-x)\delta_{ij},$
i.e. we assume that all pairs of photons have equal overlap with each
other, which we can then set to be real without loss of generality.
We will show in the Supplemental Material that it is an appropriate
appropriate parametrization of $S$ for our experimental data. Note
that in this parametrization $P(x)$ is a polynomial of degree $n$
in $x$ \cite{Rohde2015}. 

A complication arises because we are sampling over the subset of collision-free
events due to the threshold nature of the detector. This means that
we must assign probability $0$ to events containing a collision,
and increase the probability of noncollision events by a factor $C,$
where $C(x)=1/\sum_{i}P_{i}(x)$ is the normalization factor obtained
by summing all non-collision events. Inconveniently, the total probability
of obtaining a non-collision event depends on the level of indistinguishability,
meaning that $C(x)$ must estimated as well. We estimate $C(x)$ using
a Monte Carlo procedure, sampling uniformly over $10^{4}$ possible
output modes. 

Figure 2 illustrates the intuition behind the ML approach for boson
sampling. The main figure shows ten arbitrarily chosen samples from
our experiment, for $n=7,$ as a function of the level of mutual indistinguishability
$x.$ It can be observed that the output probability for each event
is a function of the level of imperfections in the system, and that
these probabilities are not monotonic in $x$. By noting which samples
we have observed, we can obtain information about the level of indistinguishability
at which our sampler is operating. 

The inset shows our Monte Carlo estimate of the fraction of collision
events (i.e. $1-C(x))$. Note that since the photon number increases
while the number of modes remains constant, the probability of a collision
event goes up strongly as a function of photon number. Having estimated
the correction factor due to collision events and computed each output
probability as a function of the level of mutual indistinguishability,
we can now maximize the likelihood function to obtain an estimate
for $x.$

Figure 3a shows the results our estimation. We compute that across
all our experiments, our photons have a mean wavefunction overlap
of $x=0.89\pm0.02,$ as shown by the black points in Fig. 3a, corresponding
to a HOM dip depth of $x^{2}=0.79$. The error bars in Fig. 3a are
given by the point where the relative likelihood is smaller than 0.05.
Fig. 3b shows the likelihood functions evaluated from $0<x<1.$ Note
that for all $n,$ $\Lambda(x=1)\gg\Lambda(x=0),$ which explains
why the effect of partial distinguishability was not detected by previous
tests, which compared only fully distinguishable and fully indistinguishable
photons \cite{Wang2019b}, or combinations of such \cite{Bentivegna2014},
as was noted previously by \cite{Dai2020}. 

Figure 3c shows an estimate of the accuracy of our method, as a function
of photon number. The variation in the size of the error bars in figure
3a) is mainly attributable to the vastly differing numbers of samples.
To get an estimate of the efficiency per sample, we numerically simulate
processing 10000 samples of each photon number at $x=1,$ and compute
the resulting uncertainty in estimating $x$ by looking at the relative
likelihood. We find that the error of our method decreases with the
number of photons. The intuition behind this is that since the indistinguishability
is a polynomial of degree $n$ in $x$ \cite{Renema2018a}, increasing
the number of photons increases the changes in probability around
the maximum of the likelihood function.

\begin{figure}
\includegraphics[width=8.8cm]{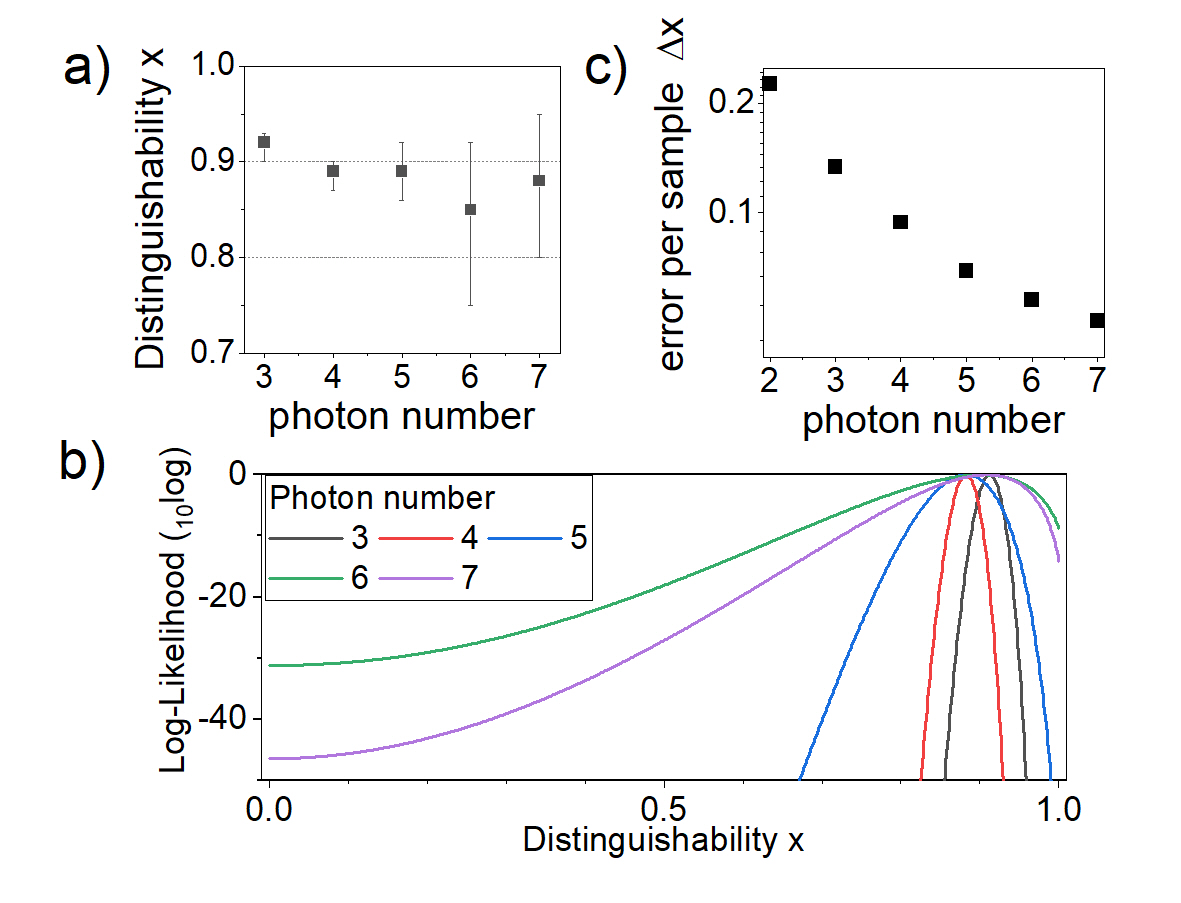}

\caption{ \emph{a)} Maximum likelihood estimation of the partial distinguishability
$x=\langle\psi_{i}|\psi_{j}\rangle,$ from a sparse series of samples.
\emph{b) }Log-likelihood functions plotted between $0\protect\leq x\protect\leq1.$
The colours indicate the number of photons. \emph{c) }numerical simulation
of the accuracy of our experiment, normalized to a constant number
of samples.}

\end{figure}

\section{Role of other imperfections}

Figure 3 raises an important question: if the wavefunction overlap
between our photons is approximately $x=0.89\pm0.02$, why does an
independent measurement of the overlap between our photons via the
Hong-Ou-Mandel effect measure $x=0.981$ \cite{Wang2017}? The answer
to this question is that imperfections in boson sampling compound
one another, resulting in a reduction of the effective photon distinguishability.
This was shown explicitly for indistinguishability and loss \cite{Renema2018b},
as well as for indistinguishability and noise on the matrix $M$ \cite{Shchesnovich2019a}.
In both of these cases, the physical picture is that noise sources
affect the interference pattern in similar ways, and can therefore
be thought of as strengthening one another. Therefore, we must look
to the other noise sources in our experiment (dark counts, multiphoton
emission, and misspecification of $M)$ to explain the discrepancy
between our inferred $x$ and that observed from HOM measurements.

To demonstrate this physical mechanism, we have performed a full numerical
simulation of our experiment, for $n=3.$ The procedure is as follows:
we generate a set of samples which contains all the same imperfections
as our real experiment, at the strengths at which we have measured
them to be present in the experiment from independent characterization
experiments. We then analyse these samples using the same procedure
which we used to analyse the actual experimental data, and report
on the observed value of $x.$

To generate these samples, we use a Markov Chain Monte Carlo sampling
routine, adapted from \cite{Neville2017}. Details are provided in
the Supplemental Material. We account for dark counts by adding some
samples with $n=2$ among all three possible combinations of pairs
of input modes, and generating the third detection event according
to the measured dark count rate of our detectors. We account for multiphoton
emission by generating samples from all nine combinations of states
containing two of the wanted photons and one noise photon. We account
for misspecification (measurement error) on the independent measurement
of the interferometer by perturbing our observed interferometer $M$
with elementwise Gaussian noise. We assess the strength of each of
these noise sources independently from separate calibration experiments
\footnote{The details of these experiments are given in the Supplemental Material}:
a $3$\% probability of a dark count, a $1.2$\% probability of double
photon emission per mode, and approximately $1\%$ error on measuring
the elements of $M.$

Table 1 shows the impact of these noise sources. By performing a series
of simulations where we switch these imperfections on one at a time,
and reconstructing $x$ at each step, we can get an idea of the cumulative
impact which each of these imperfections has on our boson sampler.
Table 1 can therefore also be interpreted as an error budget for our
boson sampler, in that it shows which imperfections most reduce the
level of indistinguishability. In particular, the effect of multiphoton
states is almost a factor 4 larger than that of misspecification of
M and of dark counts, but of the same order as the reduction of $x$
due to true indistinguishability. This shows that the main challenges
in improving the quantumness of our boson sampler lie in improving
the quantum dot source. 

\begin{table}
\begin{tabular}{|c|c|c|}
\hline 
\emph{Imperfection} & \emph{Value of $x$ inferred} & \emph{$_{10}$}log\emph{$\left(\Lambda_{s}/\Lambda_{e}\right)$ }\tabularnewline
\hline 
\hline 
$x$ from HOM experiment & 0.981 $\pm$ 0.002 & -130\tabularnewline
\hline 
Misspecification of $M$ & 0.970 $\pm$ 0.008 & -83\tabularnewline
\hline 
Multiphoton states & 0.924 $\pm$ 0.009 & -2.5\tabularnewline
\hline 
Dark counts & 0.913 $\pm$ 0.010 & 0\tabularnewline
\hline 
\end{tabular}

\caption{Error budget for a boson sampler. The first column shows the value
of the indistinguishability when that particular imperfection and
the ones above it in the table are applied. The second column shows
the relative likelihood of the corresponding value of $x$ as derived
from the simulation, normalized to the maximum value of the likelihood
function derived from the experimental data.}
\end{table}

To compare our simulation to our experimental data, we also report
in Table 1 the relative likelihood of the value of $x$ (expressed
as its 10-log) from our simulations ($\Lambda_{s})$ relative to the
likelihood derived from our expeirmental data ($\Lambda_{e})$. This
shows that when we `switch on' all imperfections, our simulation predicts
a value of $x$ within the error bar of the one which we measure in
our experiment, and hence that our simulation and our measurement
are consistent. This validates the picture of our inferred maximum
likelihood indistinguishability value being essentially an effective
value that takes into account all other known sources of error in
the experiment. 

\section{Real-time monitoring}

A practical application of our result is that we can monitor the strength
of the noise sources in our experiment in real time, by making use
of a rolling average. In this way, we can detect changes in the noise
parameters of our experiment in real time, opening up a valuable diagnostic
tool for stabilizing an experimental boson sampling setup. 

To illustrate this procedure, Figure 4 shows a case in which we have
artificially increased the $g^{(2)}$ of our photon source by increasing
the excitation power of the pump pulse from $1.19\times10^{-17}$
mJ to $4.58\times10^{-17}$ mJ. To obtain the $i$-th point in this
graph, we compute the maximum likelihood estimate of the $i$-th to
$i+10000$-th consecutive samples from the experiment. The resulting
estimate shows a clear continuous decrease in the effective photon
indistinguishability, demonstrating the use of this method for real-time
monitoring.

\begin{figure}
\includegraphics[width=8.8cm]{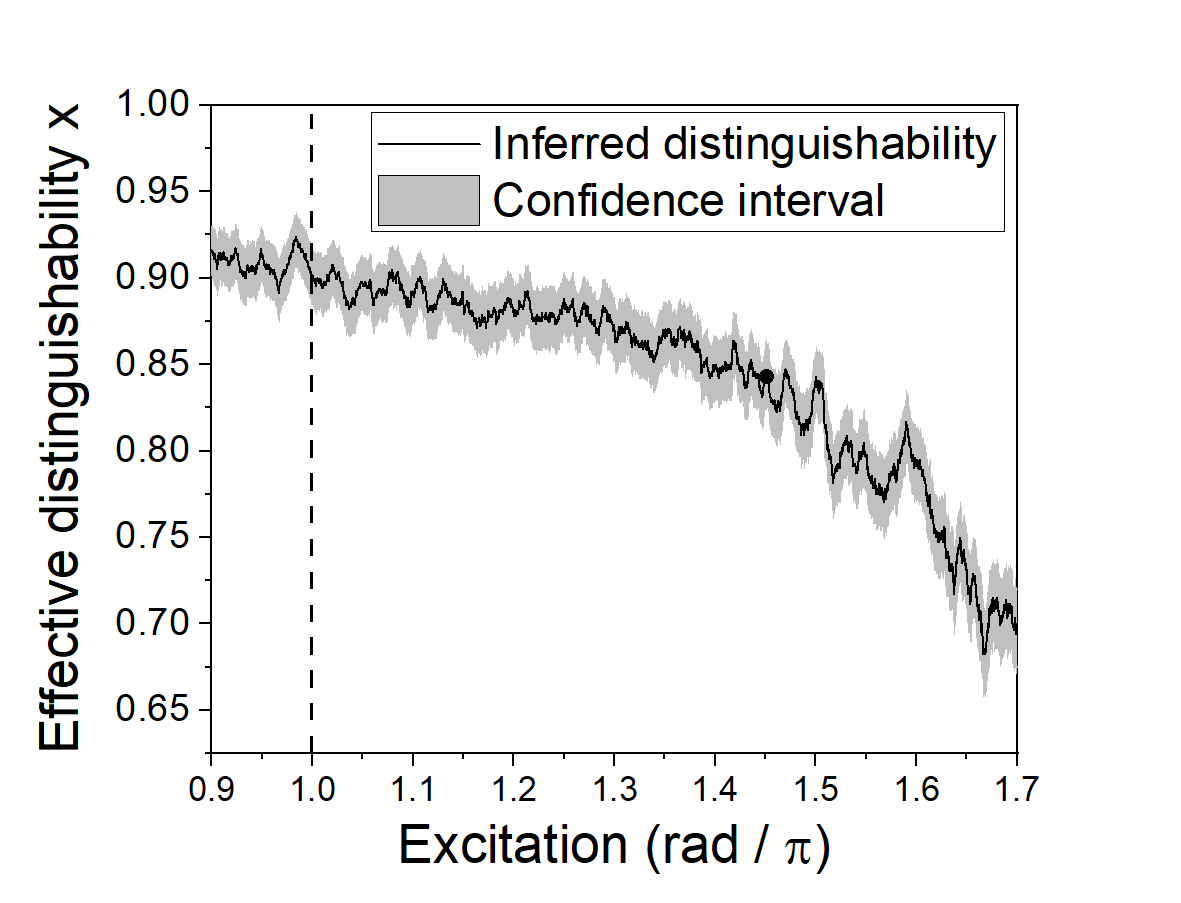}

\caption{Real time monitoring of the effective degree of photon indistinguishability
in a boson sampler as a function of excitation power. To decrease
the effective degree of indistinguishability, the pump power on the
quantum dot single photon source was increased to push the exitation
away from the ideal $\pi$-pulse (indicated with a dashed line). The
black curve shows a rolling avergae estimate of $x,$ averaging over
10.000 consecutive samples, with the grey band showing the 95\% confidence
interval. }

\end{figure}

\section{Discussion \& Conclusion}

An important restriction in our approach is that Equation 1 is expensive
to evaluate, since it requires the computation of $n!$ many permanents,
each of which comes at a cost of $n2^{n}$ computational steps. This
can be reduced to $2n2^{2n}$ by applying a multidimensional extension
to Ryser's formula \cite{Tichypartial}, but this still restricts
application of this approach to approximately 25 photons \cite{Wusupercomputer}.
Two approaches are possible: either to use the approximate probabilities
of \cite{Renema2018a,Renema2018b,Shchesnovich2019a,Shchesnovich2019b,Moylett2019,Renema2019},
or to use simpler circuits for the measurement of $x,$ as was done
in the benchmarking of Google's random circuit sampling experiment
\cite{Arute2019}. We leave the issue of finding the appropriate circuits
for the photonic case as an open problem for future work. Another
further extension of this work would be to strengthen our belief in
the thresholds for classicality by teing them to a complexity conjecture,
as was done for random circuit sampling \cite{Aaronson2019}. 

In conclusion, we have demonstrated how to infer the quality of a
boson sampler from a sparse series of samples. We have demonstrated
that the measured indistinguishability is an effective value, which
accounts for a series of imperfections. These results show how to
demonstrate that a candidate quantum advantage demonstration using
photonics outperforms the best known simulation algorithms. 
\begin{acknowledgments}
J.J.R. acknowledges NWO Veni. We thank Reinier v.d. Meer, Pim Venderbosch,
Lars Corbijn van Willenswaard, Chris Toebes and Pepijn Pinkse for
discussions.
\end{acknowledgments}

\end{document}